\documentclass[aps,prl,showpacs,twocolumn,superscriptaddress]{revtex4-2}
\usepackage[english]{babel}
\usepackage[T2A]{fontenc}
\usepackage[koi8-r]{inputenc}
\usepackage{amssymb,amsmath,amsfonts,amsthm}
\usepackage{mathrsfs}
\usepackage[mathcal]{eucal}

\usepackage[dvips]{graphicx}
\usepackage[colorlinks]{hyperref}

\usepackage{chngcntr}

\begin{document}

\title{Effect of  thermal fluctuations on the nontrivial topology of the $d+id$  superconducting  phase}

\author
{A.G.~Groshev}
\email{groshev_a.g@mail.ru}
\address{The Udmurt Federal Research Center, T. Baramzinoy St. 34, Izhevsk 426067, Russia}
\author
{A.K.~Arzhnikov}
\email{arzhnikof@bk.ru}
\affiliation{The Udmurt Federal Research Center, T. Baramzinoy St. 34, Izhevsk 426067, Russia}


\date{\today}

\begin{abstract}

The behavior of the topological index, characterizing  the properties of superconducting 
phases of quasi-two-dimensional systems with nontrivial topology, is investigated depending 
on the temperature and parameters of the effective non-Hermitian Hamiltonian. For this purpose, 
a method of calculating the topological index, based on a self-consistent functional-integral 
theory, is proposed.
The method makes it possible to take into account thermal fluctuations and study the behavior 
of the topological index as a function of temperature and Hamiltonian parameters. The chiral 
$d+id$  superconducting phase of a quasi-two-dimensional model with effective attraction 
between the electrons located at the nearest sites of a triangular lattice is considered. It 
is shown that the characteristic features in the energy dependence of the self-energy part, 
which arise when thermal fluctuations are taken into account, have a structure that does not 
lead to a change in the topological properties of the system. It is found that thermal 
fluctuations, as well as an increase in effective attraction in this system, contribute to 
the expansion of the temperature region, in which the value of the topological index is close 
to the integer $C_1\simeq-2$. 
\end{abstract}


\maketitle

\section{Introduction}

In recent decades, a field of research related to the nontrivial topology of electronic states has been actively 
developing in condensed matter physics. It is obvious that, in addition to the unusual fundamental properties of 
quasiparticles in these phases, the interest of researchers is attracted by the prospects for their technical use 
in fault-tolerant quantum computers, in the implementation of high-speed information transfer, and in spintronics 
(see, for example, \cite{Valkov_2022,Nayak_2008,Zeng_2018}). In real experiments, systems are under the influence 
of external and internal flows of energies and particles, which violates the conservation laws in closed systems 
and makes it impossible to describe the systems under consideration by wave functions in Hilbert space, which 
corresponds to the ground state of the system. This forces one to consider effective non-Hermitian Hamiltonians, 
which assume the damping of quantum states. Naturally, the question arises as to whether new topological properties 
can arise in non-Hermitian systems. Considerable attention has been paid to these issues in recent years (see, for 
example, \cite{Yuto_2020,Kawabata_2019,Long_2022,Markov_2021}). 

The main purpose of this work is to study the effect of thermal fluctuations on the topological index (TI). The paper 
considers a quasi-two-dimensional model system of a superconductor with  effective attraction of electrons located at 
neighboring sites of a triangular lattice. It should be noted that although the work is not aimed at describing real 
superconductors, the considered Hamiltonian and the chosen parameters were used by us earlier to describe layered 
compounds with a triangular lattice, such as $Na_{x}CoO_{2}yH_{2}O$ sodium cobaltites intercalated with water 
\cite{Groshev_2021}. It is important that, in such a model, superconductivity with a nontrivial topology arises in a 
natural way \cite{Zhou_2008}.

The topology of the superconducting phase in quasi-two-dimensional materials is characterized by an integer value of 
the topological index TI which describes the nonlocal characteristics of the many-particle wave function of the electronic 
ensemble and is expressed in terms of one-electron Green functions \cite{Valkov_2019} 
\begin{equation}
\begin{array}{c}
\label{eq:Ti}
\displaystyle
C_{1}=\frac{\varepsilon_{\mu\nu\gamma }}{24\pi^{2}}\int\limits_{-\infty}^{+\infty}d\omega
\int\limits_{BZ} d^{2}k\times
\\
\times tr\left[G\partial_{\mu}G^{-1}G\partial_{\nu}G^{-1}
G\partial_{\gamma}G^{-1}\right],
\end{array}
\end{equation}
where $\varepsilon_{\mu\nu\gamma }$ is the antisymmetric Levi-Civita tensor, (summation is assumed over repeated indices 
$\mu, \nu, \gamma $ in (\ref{eq:Ti}), $G(i\omega)$ is the Matsubara Green function, $\mu,\nu,\gamma $ are the frequency-momentum 
indices $(k_{1},k_{2}$, and $BZ$ is the Brillouin zone. The considered TI is also used to characterize quasi-two-dimensional 
topological insulators, integer and fractional quantum Hall effects \cite{Ishikawa_1987}. When considering the Green functions 
of a superconductor in the mean field approximation \cite{Rachel_2010,Raghu_2008} it is possible to integrate over frequency in 
(\ref{eq:Ti}). In this case, the expression for TI reduces to the well-known definition of the Chern number associated 
with the Berry phase in momentum space \cite{Niu_1985,Valkov_2019}. Accounting for many-particle effects, disorder, and fluctuations 
leads to the energy dependence of the self-energy part of the one-electron Green function. In some cases, this dependence can affect 
the topological phase or induce transitions to new topological phases \cite{Wang_2011}. In Ref. \cite{Budich_2012}, the role of 
quantum fluctuations in topological phase transitions with spin and anomalous Hall effects was studied. It was shown that dynamic 
fluctuations bring a topologically trivial insulator into the phase with an integer Chern number. In Ref. \cite{Zheng_2019}, a topological 
phase transition was studied in a two-dimensional disordered system with effective repulsion between electrons located at the same site. 
It was found that the repulsive interaction of electrons at a site contributes to the preservation of the topological phase in the 
disordered system. In \cite{Budich_2012}, a general mechanism was proposed that ensures the appearance of a topological phase 
transition. This mechanism is due to the divergences in self-energy. According to Ref. \cite{Groshev_2022}, singularities in the 
energy dependence of the self-energy part, can also arise as a result of resonant scattering of charge carriers by thermal fluctuations 
of electron-hole pairs, which significantly renormalizes the energy dependence of the self-energy. In this paper, to calculate TI, 
taking into account thermal fluctuations in superconducting phases, the functional-integral method is used in the general calculation 
scheme, which is similar to the calculation of TI in the quantum Hall effect based on the Kubo formula \cite{Thouless_1982}.

\section{Calculation of the topological index taking into account thermal fluctuations }

As is known, the topological index in the integer quantum Hall effect is proportional to the Hall conductivity that can be calculated 
by the Kubo formula \cite{Mahan_2000}. In the case of superconducting phases with nontrivial topology, we also use this approach. In 
the functional-integral theory the Kubo formula for conductivity has the following form: 
\begin{equation}
\label{eq:Cubo}
\displaystyle
\sigma_{\mu\nu}=\lim_{\omega\rightarrow 0}\frac{1}{\Omega_{n}}\langle
\left[\Pi_{\mu\nu}(0)-\Pi_{\mu\nu}(i\Omega_{n})\right]\rangle
\mid_{i\Omega_{n}=\omega+i0},
\end{equation}
where the angle brackets $\langle ...\rangle$ denote averaging over fluctuating fields, $\Omega_{n}=2\pi nT $ is the boson Matsubara 
frequency, $\Pi_{\mu\nu}(i\Omega_{n})$ is the Matsubara current-to-current correlation function 
\begin{equation}
\displaystyle
\Pi_{\mu\nu}(i\Omega_{n})=\int\limits_{0}^{\beta}d\tau\langle T_{\tau}J_{\mu}(\tau)J_{\nu}(0)
\rangle\exp(i\Omega_{n}\tau),
\end{equation} 
where $J_{\mu(\nu)}$ are components of the current operators, $T_{\tau}$ is the ordering operator in "imaginary time". 
The angle brackets denote quantum statistical averaging. As is known, fluctuating fields in the functiona-integral theory for 
superconducting phases arise as a result of the Hubbard-Stratonovich transformation, which allows one to replace the many-particle 
problem of interacting electron-hole pairs by the one-particle problem with an electron-hole pair interacting with auxiliary random 
fields \cite{Groshev_2020,Groshev_2021} (see Appendix A).

To calculate TI, we restrict ourselves to the one-loop approximation, which allows one to obtain a standard expression for TI at 
$T=0$ and, at the same time, take into account thermal fluctuations at finite temperatures. In this approximation, the correlation 
function $\Pi_{\mu\nu}(i\Omega_{n})$  is written as follows: 
\begin{equation}
\label{eq:Current-current2}
\displaystyle
\Pi_{\mu\nu}(i\Omega_{n})=-\frac{1}{\beta}\sum_{m}
Sp\left[J_{\nu}F(i\omega_{m}-i\Omega_{n})J_{\mu}F(i\omega_{m})\right],
\end{equation}
where $F(i\omega_{m})$ $J_{\nu}$ and $J_{\nu}$ are the Matsubara Green function and component of the current operator averaged over 
fluctuating fields, $\omega_{n}=\pi T(2m+1)$ are the Matsubara frequencies for Fermi particles, $SpA=\sum_{k}trA(k)$, $tr$ implies 
summation over spin variables. In deriving this expression, it was taken into account that $\omega_{m}+\Omega_{n}=\pi T(2m+1+2n)=\omega_{n'}$.  
Note that the potential of fluctuating fields $\Delta{\cal U}_{j\delta}$ in the functional-integral theory is chosen so that its average 
value $\langle \Delta{\cal U}_{j\delta}\rangle =0$ [22]. Therefore, the components of the current operators $J_{\nu}$ averaged over 
fluctuating fields are expressed in terms of the Matsubara Green function with the averaged order parameter 
\begin{equation}
\displaystyle
J_{\nu}=e\partial_{\nu}\langle{\cal H}\rangle =e\partial_{\nu}{\cal H}_{AV}=
-e\partial_{\nu}\left(G^{AV}\right)^{-1},
\end{equation}
where $\partial_{\nu}=\partial/\partial_{k_{\nu}}$. At $T\ne 0$ there arise thermal fluctuations. Therefore, when obtaining an 
expression for TI, it is necessary to take into account the discreteness of the Matsubara frequencies $\omega_{n}$. To this end, the 
residue theorem is used and the summation over Matsubara frequencies is replaced by integration along the cut lines $z=E+i0+i\Omega_{n}$, 
$z=E-i0+i\Omega_{n}$,  and $z=E+i0$, bypassing the point $z=i\Omega_{n}$ in the upper half-plane, and along the line $z=E-i0$ in the 
lower half-plane of the complex energy. By this means we can go to the retarded (advanced) Green functions 
$F^{R(A)}(E)=\langle (E-{\cal H}\pm i0)^{-1}\rangle$. As a result, for the correlation function (\ref{eq:Current-current2}) we obtain  
\begin{equation}
\label{eq:Current-current3}
\begin{array}{c}
\displaystyle
\Pi_{\mu\nu}(i\Omega_{n})=\frac{1}{4\pi i}\int\limits_{-\infty}^{+\infty}dE th\left(\frac{\beta E}{2}\right)
\times
\\
\times Sp\left\{J_{\nu}\left[F^{A}(E)-F^{R}(E)\right]J_{\mu}F^{R}(E+i\Omega_{n})+
\right.
\\
\left.
J_{\nu}F^{A}(E-i\Omega_{n})J_{\mu}\left[F^{A}(E)-F^{R}(E)\right]\right\},
\end{array}
\end{equation}
where $th\left(z\right)$ is the hyperbolic tangent. Restricting ourselves to the quasi-two-dimensional case and passing in 
(\ref{eq:Cubo}) to the limit $\Omega_{n}\to 0$, we single out the antisymmetric part, which determines the Hall conductivity 
$\sigma_{H}=\frac{e^2}{2\pi\hbar}C_{1}$ in the integer quantum Hall effect. As a result, we obtain a generalizing expression 
for TI, taking into account thermal fluctuations 
\begin{equation}
\label{eq:Sigma_H_T}
\begin{array}{c}
\displaystyle
C_{1}=\int\limits_{-\infty}^{+\infty}dE
\int\limits_{-\pi}^{\pi}\int\limits_{-\pi}^{\pi} \frac{dk_{1}dk_{2}}{16\pi^{2}}
th\left(\frac{\beta E}{2}\right)\times
\\
\displaystyle
\times 
tr\left[\partial_{k_{1}}G^{-1}_{AV}(E)K^{-}(E)\partial_{k_{2}}G^{-1}_{AV}(E)\partial_{E}K^{+}(E)
-
\right.
\\
\displaystyle
\left.
-K^{-}(E)\partial_{k_{1}}G^{-1}_{AV}(E)\partial_{E}K^{+}(E)\partial_{k_{2}}G^{-1}_{AV}(E)\right].
\end{array}
\end{equation} 
Here we introduce the notation  $K^{+}(E)=$ $[F^{A}(E)+F^{R}(E)]/2$, $K^{-}(E)=[F^{A}(E)-F^{R}(E)]/2$, 
$\partial_{E}=\partial/\partial_{E}$, $k_{1}$ and are the main reciprocal lattice vectors. To obtain an explicit expression 
for TI of a quasi-two-dimensional system, it is necessary to multiply the matrices under the sign $tr$ in (\ref{eq:Sigma_H_T}). 
To do this, it is convenient to use their expansion in Pauli matrices (see Appendix B). It should be noted that passing from 
summation over the Matsubara frequencies $\omega_{n}$ to integration along the imaginary axis 
$\frac{1}{\beta}\sum_{m}\to\int\frac{d\omega}{2\pi}$, valid for $T\to 0$, in the limit $\Omega_{n}\to 0$ from (\ref{eq:Cubo}) 
we get the standard expression for TI (\ref{eq:Ti}). In contrast to (\ref{eq:Ti}), integration over energy in 
(\ref{eq:Sigma_H_T}) is carried out along the real axis.

\section{ Model and results}

In this paper we consider a quasi-two-dimensional model system with effective attraction of electrons at neighboring sites 
of a triangular lattice with the Hamiltonian  
\begin{equation}
\label{eq:hamiltonian1}
{\hat {\cal H}}=\sum_{i,j,s}t_{ij}{\hat c}_{is}^{+}{\hat c}_{js}^{}-
\sum_{j}\mu{\hat n}_{j}-
V\sum_{j,\delta}{\hat n}_{j\uparrow}{\hat n}_{j+\delta\downarrow},
\end{equation} 
where $t_{ij}=-t$ are the matrix elements of electron hopping to the nearest sites; $\hat{c}_{js}^{+}(\hat{c}_{js})$ are the 
operators of creation (annihilation) of an electron at site $j$ with spin projection $s$; $n_{js}=\hat{c}_{js}^{+}\cdot\hat{c}_{js}$   
is the operator of the number of electrons at site $j$ with spin projection $s$; $n_{j}$ is the operator of the total number of 
electrons at site $j$; $\mu$ is the chemical potential; $V$ is the parameter of inter-electron attraction. The Hamiltonian of the 
system considered with the averaged order parameter ${\hat {\cal H}}_{AV}$ in the quasi momentum representation has the form (see 
\cite{Groshev_2021}): 
\begin{equation}
\label{eq:Nambu_k}
\begin{array}{c}
\displaystyle
\displaystyle
{\cal H}_{AV}(k)=
\left[
\begin{array}{cccc}
{\cal H}_{AV}^{\uparrow\uparrow}(k)&
{\cal H}_{AV}^{\uparrow\downarrow}(k)\\
{\cal H}_{AV}^{\downarrow\uparrow}(k)&
{\cal H}_{AV}^{\downarrow\downarrow}(k)
\end{array}
\right],
\\
\displaystyle
{\cal H}_{AV}^{\uparrow\uparrow}(k)=\varepsilon_{k},\qquad 
{\cal H}_{AV}^{\downarrow\downarrow}(k)=-\varepsilon_{k}
\\
\displaystyle
{\cal H}_{AV}^{\uparrow\downarrow}(k)=-2V\overline{\Delta}V_{k},
\\ 
\displaystyle
{\cal H}_{AV}^{\downarrow\uparrow}(k)=\left({\cal H}_{AV}^{\uparrow\downarrow}(k)\right)^{*},
\\
\varepsilon_{k}=-2t\left[\cos{k_{1}}+\cos{k_{2}}
+\cos{\left(k_{2}-k_{1}\right)}\right]-\mu,
\\
\displaystyle
V_{k}(\alpha)=\cos{k_{1}}+\exp(i\alpha)\cos{k_{2}}+
\\
\displaystyle
+\exp(-i\alpha)\cos{(k_{2}-k_{1})},
\end{array}
\end{equation}
where $\varepsilon_{k}$ is the dispersion law of electron energy on a triangular lattice with hoppings within the first 
coordination sphere; $V_{k}(\alpha)$ is the dispersion law of the superconducting order parameter with symmetry determined 
by the phase value $\alpha$. The retarded (advanced) effective Green functions $F^{R(A)}(E)=(E-{\cal H}_{AV}-\Sigma^{R(A)}(E))^{-1}$ 
entering into (\ref{eq:Sigma_H_T}) have the form: 
\begin{equation}
\label{eq:F_k}
\begin{array}{c}
\displaystyle
F(E)=
\left[
\begin{array}{cccc}
F^{\uparrow}(E)&
F^{\uparrow\downarrow}(E)\\
F^{\downarrow\uparrow}(E)&
F^{\downarrow}(E)
\end{array}
\right],
\\
\displaystyle
F^{\uparrow (\downarrow)}(E)=\frac{E\pm\varepsilon_{k}-\Sigma^{\downarrow(\uparrow)}_{}(k,E)}
{\left[E-E_{+}(k)\right]\left[E-E_{-}(k)\right]},
\\
\displaystyle
F^{\uparrow\downarrow (\downarrow\uparrow)}(E)=
\frac{\Sigma^{\uparrow\downarrow (\downarrow\uparrow)}_{}(k,E)}
{\left[E-E_{+}(k)\right]\left[E-E_{-}(k)\right]},
\\
\displaystyle
E_{\pm}(k)=\left[\Sigma^{\uparrow}_{}(k,E)+\Sigma^{\downarrow}_{}(k,E)\right]/2\pm
\\
\displaystyle
\pm\left[\left(\varepsilon_{k}+\left[\Sigma^{\uparrow}_{}(k,E)-
\Sigma^{\downarrow}_{}(k,E)\right]/2
\right)^{2}+
\right.
\\
\left.
\Sigma^{\uparrow\downarrow}_{}(k,E)\Sigma^{\downarrow\uparrow}_{}(k,E)\right]^{1/2}.
\end{array}
\end{equation} 
For simplicity, we do not indicate here the indices  in the notation of the retarded (advanced) Green functions. 
Just as in \cite{Groshev_2022}, we restrict ourselves to an approximation quadratic in  fluctuating potential 
for the self-energy part. The components of the self-energy part have poles at the boundaries of the energy gap. 
These anomalies arise as a result of resonant scattering of charge carriers by thermal fluctuations of the 
electron-hole pairs \cite{Groshev_2022,Izyumov_1965}. However, the resulting pole structure 
$\Sigma(k,E)\propto (E-E_{k})^{-1}+(E+E_{k})^{-1}$ cannot give rise to new topological phases and transitions 
\cite{Wang_2011,Wang_2012}. In addition, the approximation we use, which is quadratic in terms of fluctuating 
potential, allows one to obtain analytical expressions for the derivatives of the components of the self-energy 
part with respect to energy that simplify numerical calculations of the topological index $C_1$. Expanding 
${\hat {\cal H}}_{AV}$ (\ref{eq:Nambu_k}) and $F^{R(A)}(E)$ (\ref{eq:F_k}) in Pauli matrices (see Appendix B), we 
obtain an explicit expression for  $C_1$ (\ref{eq:Sigma_H_T}) (see Appendix C). 

Calculations of the topological index  $C_1$ (\ref{eq:Sigma_H_T}) were carried out in the range of concentrations 
where the superconducting phase of the system under consideration has a $d+id$ type of symmetry ($\alpha=2\pi/3$) 
(see the phase diagram in Fig.~2) at a value of the interelectronic attraction parameter $V=t$ typical of 
$Na_{x}CoO_{2}yH_{2}O$ compounds \cite{Groshev_2021}. In the concentration range where the superconducting phase has 
a generalized $s$ symmetry ($\alpha=0$), from the explicit expression for $C_1$ (\ref{eq:Sigma_H_T}) it follows that  
$C_1=0$ for any parameters (see Appendix C). In Ref. \cite{Groshev_2022} it was shown that, as a result of resonant 
scattering of charge carriers on thermal fluctuations of electron-hole pairs, in the normal and pairing components 
of the self-energy part there arise singularities (poles) at energy values corresponding to the boundaries of the 
energy gap $E\simeq\pm E_k$.

The results of calculating the temperature dependence of the topological index $C_1$ for $d+id$ pairing with account 
for the above features of the self-energy part, are shown in Fig.~1 for three values of the effective attraction constant:
$V=0.5$, $V=1$ and $V=2$. For comparison, is also presented the TI temperature dependence calculated in the Hartree-Fock 
(HF) approximation without allowance for thermal fluctuations. It can be seen that when fluctuations are taken into 
account, the temperature range in which the TI value is close to the integer $C_1\simeq-2$ becomes much wider than in 
the HF approximation. Moreover, in contrast to the results of the HF approximation, when thermal fluctuations are taken 
into account, with increasing $V$, the region with $C_1\simeq-2$ expands and approaches the superconducting transition 
temperature $T_{c}$ at $V=2$. Assuming that thermal fluctuations play the role of disorder, the latter result is 
consistent with that of Ref.\cite{Zheng_2019}, in which the repulsive interaction between electrons located at the same 
site leads to the expansion of the topological phase region in a disordered system.
\begin{figure}[h]
\begin{center}
\includegraphics[width=\linewidth]{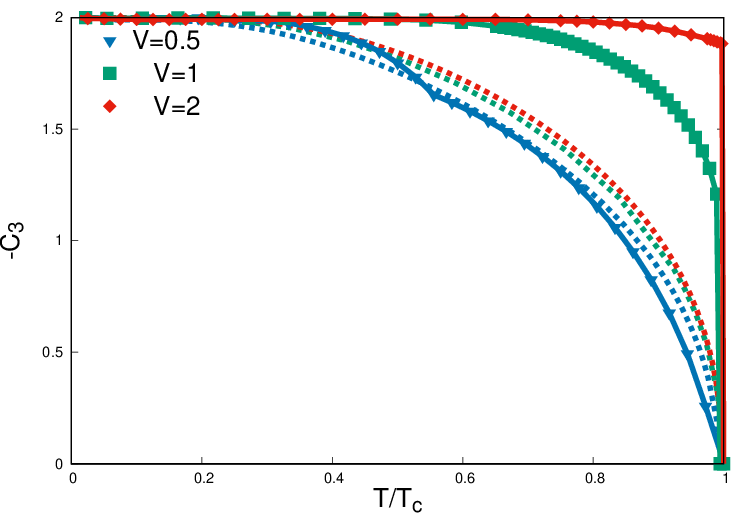}
\caption{Temperature dependence of the topological index $C_1$ for $d+id$ pairing at three values of the constant of 
effective attraction $V$ and electron concentration $n\simeq 1.5$. The dotted lines represent the temperature dependences 
of $C_1$ in the Hartree-Fock approximation for the same parameter values.}
\end{center}
\end{figure} 
In Figure~3 the temperature dependence of the topological index $C_1$ for $d+id$ pairing is presented at three charge 
carrier concentrations: $n\simeq1.5$, $n\simeq1.6$ and $n\simeq1.7$. From the phase diagram in Fig.~2 it is seen that, 
when deviating from the optimal doping $n\simeq1.5$, the temperature of the superconducting transition $T_{c}$ decreases 
sharply. Despite this, the temperature range in units $T_{c}$, with a TI close to the integer value $C_1\simeq-2$ changes 
in Fig.~3 insignificantly. 
\begin{figure}[h]
\begin{center}
\includegraphics[width=\linewidth]{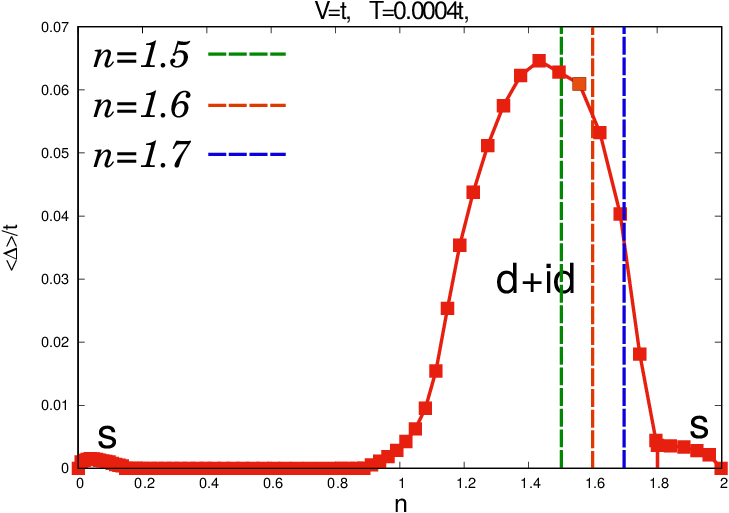}
\caption{The dependence of the amplitude of the averaged order parameter on the charge carrier concentration at 
a temperature $T=0.0004t$ (phase diagram).}
\end{center}
\end{figure}
\begin{figure}[h]
\begin{center}
\includegraphics[width=\linewidth]{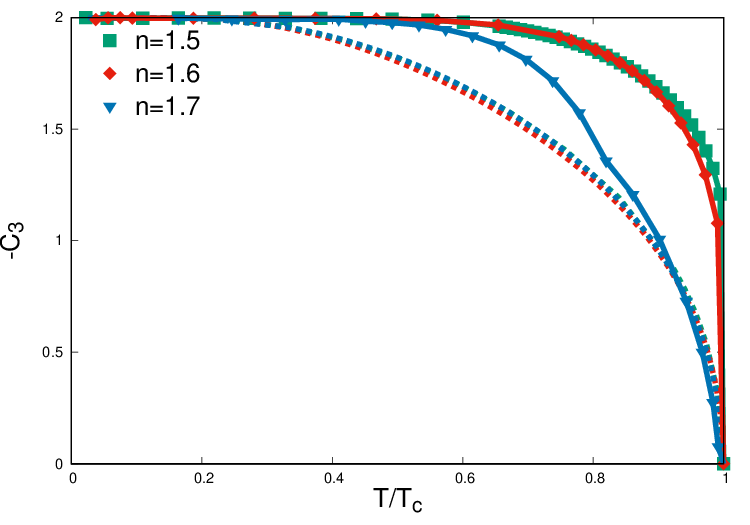}
\caption{ Temperature dependence of the topological index $C_1$ for $d+id$ pairing at three values of the electron 
concentration $n$ and the constant of effective attraction V=1. The dotted lines represent the temperature dependences of 
$C_1$ in the Hartree-Fock approximation for the same parameter values.}
\end{center}
\end{figure}
Figure~4 presents the dependence of TI on the electron concentration at temperatures $T/T^{*}_{c}=0.02$, 
$T/T^{*}_{c}=0.2$ and $T/T^{*}_{c}=0.8$. Since the temperature of the transition to the superconducting 
state $T_{c}$ depends on the concentration of charge carriers $n$ (see Fig.~2), its largest value at the 
optimal doping $n\simeq 1.5$ is chosen as  $T^{*}_{c}$ in Fig.~4. It demonstrates that, just as in the 
integer quantum Hall effect, with increasing temperature, the concentration region where the TI value is 
close to the integer value $C_1\simeq-2$ (the plateau region in the integer quantum Hall effect) decreases.  
\begin{figure}[h]
\begin{center}
\includegraphics[width=\linewidth]{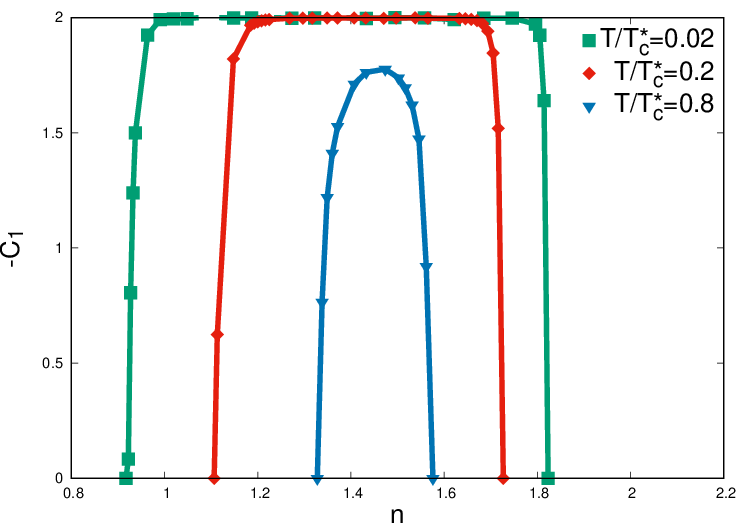}
\caption{The dependence of the topological index $C_1$ on the electron concentration for $d+id$- pairing at three temperature 
values: $T/T^{*}_{c}=0.02$, $T/T^{*}_{c}=0.2$ and $T/T^{*}_{c}=0.8$, and the constant of effective attraction V=1.}
\end{center}
\end{figure}

\section{Conclusion}

An expression for the topological index (TI) of a quasi-two-dimensional model of superconductor, which takes into account 
thermal fluctuations, is obtained in the framework of the self-consistent functional-integral theory, similarly to the calculation 
of TI in the integer quantum Hall effect. Using  the expression obtained, the TI behavior in the chiral $d+id$ superconducting 
phase of the quasi-two-dimensional one-band model is analyzed. At a temperature $T>0$, a nonzero density of states appears in 
the superconducting gap. In this case, it seems incorrect to identify the state of the system as a superconducting phase with 
nontrivial topology. However, as shown by the experiments, the system properties, associated with the nontrivial topology, are 
preserved in a certain temperature range. In particular, the quantum Hall effect takes place at nonzero temperatures when the 
inequality $\sigma_{xx}/\sigma_{xy}\ll 1$ is satisfied. In this connection, we believe that in our calculations, when the TI value 
is close to the integer value $C_1\simeq-2$, the system retains properties that are determined by the nontrivial topology of the 
chiral $d+id$ superconducting phase in the ground state. The singularities in the energy dependences of the normal and anomalous 
components of the self-energy part of the single-particle Green function, which arise as a result of resonant scattering of charge 
carriers on thermal fluctuations of electron-hole pairs, are taken into account. It is shown that these singularities do not lead 
to abrupt changes in TI, which may be interpreted as the absence of topological transitions. This is explained by the type of pole 
structure of the self-energy part of the model considered. It has been found that taking into account thermal fluctuations, as well 
as increasing the effective attraction between electrons located at neighboring sites, significantly expand the temperature region 
in which the TI value is close to the integer $C_1\simeq-2$. An expansion of this region, with thermal fluctuations taken into account, 
is also observed when the system deviates from the optimal doping level. The proposed method can be extended to the impurity disorder and 
used to self-consistently allow for the effect of thermal fluctuations on TI of topological insulators.

\acknowledgments{ This study was supported by the financing program BB-2021-121030100005-1}

\appendix*
\section{APPENDIX A: FUNCTIONAL-INTEGRAL METHOD}
\setcounter{section}{0}
\setcounter{equation}{0}
\renewcommand{\thesection}{A}
\renewcommand{\theequation}{A}
\numberwithin{equation}{section}

The functional-integral method is based on the Hubbard-Stratonovich transformation. We use this transformation 
in a two-field representation (see \cite{Izyumov_1988,Groshev_2020}), in which the amplitude $\Delta_{j,\delta}$ 
and the phase $\phi_{j,\delta}$ of the superconducting order parameter act as fluctuating fields. In this approach, 
the problem of calculating the partition function of interacting electron pairs is reduced to the problem of 
calculating the partition function of independent electron pairs located in the extended space of auxiliary fluctuating 
fields $\Delta_{j,\delta}$ and $\phi_{j,\delta}$. In the static approximation, fluctuating fields do not 
depend on time, and the partition function has the form
\begin{equation}
\label{eq:hamiltonian4}
\begin{array}{c}
\displaystyle
Z=\prod_{j,\delta}\int_{0}^{\infty}\Delta_{j,\delta }d\Delta_{j,\delta }
\int_{-\pi}^{\pi}d\phi_{j,\delta}
\exp{\left[-\beta\Omega(\Delta, \phi)\right]}, 
\\
\displaystyle
\Omega(\Delta, \phi)=\Omega^{0}(\Delta, \phi)+
\Omega^{*}(\Delta,\phi), 
\\
\displaystyle
\Omega^{0}(\Delta,\phi)=V\sum_{j,\delta}\Delta_{j,\delta}^{2},
\\
\displaystyle
\Omega^{*}(\Delta ,\phi)=-\frac{1}{\beta} \ln{Z^{*}(\Delta ,\phi)},
\\
\displaystyle
Z^{*}(\Delta ,\phi)=SpT_{\tau}\exp{\left[-\int_{0}^{\beta}d\tau 
{\hat {\cal H}}(\Delta, \phi,\tau)\right]},
\\
\displaystyle
{\hat {\cal H}}(\Delta, \phi,\tau)={\hat {\cal H}}_{HF}+
\Delta{\hat {\cal U}}(\Delta, \phi,\tau),
\end{array}
\end{equation}
where $Sp$ is the total quantum mechanical trace; $T_{\tau}$ is the operator of ordering in imaginary time 
$\tau\in[0,\beta]$, $\beta=1/k_{B}T$; ${\hat {\cal H}}_{HF}$ is the Hamiltonian of the model considered in 
the HF approximation; $\Delta{\hat {\cal U}}$ is the potential of fluctuating fields. The condition for the 
minimum thermodynamic potential together with the equations for the chemical potential, the distribution 
function of phase fluctuations, and the self-energy part of the one-electron Green function form a self-consistent 
system of equations  \cite{Groshev_2021,Groshev_2020}. The numerical solution of this system of equations 
makes it possible to determine the superconducting properties of the system under consideration, taking into 
account thermal fluctuations.

\appendix*
\section{APPENDIX B: DECOMPOSITION IN PAULI MATRICES}
\setcounter{section}{0}
\setcounter{equation}{0}
\renewcommand{\thesection}{B}
\renewcommand{\theequation}{B}
\numberwithin{equation}{section}

Introduce the following notation for matrices in (\ref{eq:Sigma_H_T}): $A=\partial_{k_{1}}G^{-1}_{AV}(E)$, 
$B=K^{-}(E)$, $C=\partial_{k_{2}}G^{-1}_{AV}(E)$ and $D=\partial_{E}K^{+}(E)$. In the two-dimensional case, 
to multiply the matrices it is convenient to use their decomposition in Pauli matrices $\sigma_{\nu}$: 
$A=A^{\nu}\sigma_{\nu}$, $B=B^{\mu}\sigma_{\mu}$ and $AB=(AB)^{\mu}\sigma_{\mu}$ 
where
\begin{equation}
\label{eq:matrix1}
\begin{array}{c}
\displaystyle
(AB)^{0}=A^{0}B^{0}+A^{x}B^{x}+A^{y}B^{y}+A^{z}B^{z},
\\
\displaystyle
(AB)^{x}=A^{x}B^{0}+A^{0}B^{x}+i\left(A^{y}B^{z}-A^{z}B^{y}\right),
\\
\displaystyle
(AB)^{y}=A^{y}B^{0}+A^{0}B^{y}+i\left(A^{z}B^{x}-A^{x}B^{z}\right),
\\
\displaystyle
(AB)^{z}=A^{z}B^{0}+A^{0}B^{z}+i\left(A^{x}B^{y}-A^{y}B^{x}\right).
\end{array}
\end{equation}
The product of matrices entering into (\ref{eq:Sigma_H_T}) has the structure $tr[ABCD-BADC]$. Since at 
$\nu\ne 0$ $tr\sigma_{\nu}=0$ and $tr\sigma_{\mu}\sigma_{\nu}=2\delta_{\mu\nu}$, from the product of matrices 
in (\ref{eq:Sigma_H_T}) there remains only the zero component $tr[ABCD-BADC]=(ABCD)^{0}-(BADC)^{0}$ which can 
be easily calculated using (\ref{eq:matrix1}): 
\begin{equation}
\label{eq:matrix2}
\begin{array}{c}
\displaystyle
tr[ABCD-BADC]=
\\
\displaystyle
=2\left[(AB)^{\nu}(CD)^{\nu}-(BA)^{\nu}(DC)^{\nu}\right].
\end{array}
\end{equation} 
As $(AB)^{0}=(BA)^{0}$, the term $\nu =0$ is absent in expression (\ref{eq:matrix2}), therefore 
\begin{equation}
\label{eq:matrix3}
\begin{array}{c}
\displaystyle
tr[ABCD-BADC]=
\\
\displaystyle
=2i\left[(A^{y}B^{z}-A^{z}B^{y})(C^{x}D^{0}+C^{0}D^{x})+
\right.
\\
\displaystyle
\left.
+(A^{x}B^{0}+A^{0}B^{x})(C^{y}D^{z}-C^{z}D^{y})+
\right.
\\
\displaystyle
\left.
+(A^{z}B^{x}-A^{x}B^{z})(C^{y}D^{0}-C^{0}D^{y})+
\right.
\\
\displaystyle
\left.
+(A^{y}B^{0}+A^{0}B^{y})(C^{z}D^{x}-C^{x}D^{z})+
\right.
\\
\displaystyle
\left.
+(A^{x}B^{y}-A^{y}B^{x})(C^{z}D^{0}-C^{0}D^{z})+
\right.
\\
\displaystyle
\left.
+(A^{z}B^{0}+A^{0}B^{z})(C^{x}D^{y}-C^{y}D^{x})\right].
\end{array}
\end{equation}

\appendix*
\section{APPENDIX C: ANALITICAL EXPRESSION FOR the TI INTEGRAND }
\setcounter{section}{0}
\setcounter{equation}{0}
\renewcommand{\thesection}{C}
\renewcommand{\theequation}{C}
\numberwithin{equation}{section}

Using the definition of matrices $A$, $B$, $C$ and $D$ (Appendix B) and the explicit form (\ref{eq:Nambu_k}) 
and (\ref{eq:F_k}, we obtain $A^{0}=0$, $C^{0}=0$,
\begin{equation}
\label{eq:B1}
\begin{array}{c}
\displaystyle
A^{x}=-2V\overline{\Delta}\left[\sin{k_{1}}-\cos{\alpha}\sin({k_{2}-k_{1})}\right],
\\
\displaystyle
A^{y}=2V\overline{\Delta}\sin{\alpha}\sin({k_{2}-k_{1})}, 
\\
\displaystyle
A^{z}=-2t\left[\sin{k_{1}}-\sin({k_{2}-k_{1}})\right],
\\
\displaystyle
C^{x}=-2V\overline{\Delta}\cos{\alpha}\left[\sin{k_{2}}+\sin({k_{2}-k_{1})}\right], 
\\
\displaystyle
C^{y}=2V\overline{\Delta}\sin{\alpha}\left[\sin{k_{2}}-\sin({k_{2}-k_{1})}\right], 
\\
\displaystyle
C^{z}=-2t\left[\sin{k_{2}}+\sin({k_{2}-k_{1}})\right], 
\end{array}
\end{equation}

\begin{equation}
\label{eq:B2}
\begin{array}{c}
\displaystyle
B^{0}=iIm\left[\frac{E-\Sigma_{1}(E)}{det(E)}\right], 
B^{x}=i2V_{1}Im\left[\frac{\Sigma^{\uparrow\downarrow}(E)}{det(E)}\right], 
\\
\displaystyle
B^{y}=i2V_{2}Im\left[\frac{\Sigma^{\uparrow\downarrow}(E)}{det(E)}\right], 
B^{z}=iIm\left[\frac{\varepsilon_{k}-\Sigma_{2}(E)}{det(E)}\right],
\end{array}
\end{equation}

\begin{equation}
\label{eq:B3}
\begin{array}{c}
\displaystyle
D^{0}=Re\left[\frac{1-\partial_{E}\Sigma_{1}(E)}{det(E)}-
\frac{\left[E-\Sigma_{1}(E)\right]\partial_{E}det(E)}{det^{2}(E)}\right],
\\
\displaystyle
D^{x}=2V_{1}Re\left[\frac{\partial_{E}\Sigma^{\uparrow\downarrow}(E)}{det(E)}-
\frac{\Sigma^{\uparrow\downarrow}(E)\partial_{E}det(E)}{det^{2}(E)}\right],
\\
\displaystyle
D^{y}=2V_{2}Re\left[\frac{\partial_{E}\Sigma^{\uparrow\downarrow}(E)}{det(E)}-
\frac{\Sigma^{\uparrow\downarrow}(E)\partial_{E}det(E)}{det^{2}(E)}\right],
\\
\displaystyle
D^{z}=-Re\left[\frac{\partial_{E}\Sigma_{2}(E)}{det(E)}+
\frac{\left[\varepsilon_{k}-\Sigma_{2}(E)\right]\partial_{E}det(E)}{det^{2}(E)}\right],
\end{array}
\end{equation}
where
\begin{equation}
\label{eq:B4}
\begin{array}{c}
\displaystyle
det(E)=det(F^{-1}(E))=\left[\frac{}{}E-E_{+}(k)\right]\left[\frac{}{}E-E_{-}(k)\right],
\\
\displaystyle
\Sigma_{1}(E)=\left[\frac{}{}\Sigma^{\uparrow}_{}(k,E)+\Sigma^{\downarrow}_{}(k,E)\right]/2,
\\
\displaystyle
\Sigma_{2}(E)=\left[\frac{}{}\Sigma^{\downarrow}_{}(k,E)-\Sigma^{\uparrow}_{}(k,E)\right]/2,
\\
\displaystyle
V_{1}=\cos{k_{1}}+\cos{\alpha}\left[\cos{k_{2}}+\cos{\left(k_{2}-k_{1}\right)}\right],
\\
\displaystyle
V_{2}=-\sin{\alpha}\left[\cos{k_{2}}-\cos\left(k_{2}-k_{1}\right)\right],\qquad\quad\quad
\\
\displaystyle
\partial_{E}det(E)=\partial_{E}det(F^{-1}(E))=\qquad\qquad\qquad\qquad\qquad\qquad
\\
\displaystyle
=\left[\frac{}{}1-\partial_{E}\Sigma_{1}(E)+\partial_{E}\Sigma_{2}(E)\right]\times
\\
\displaystyle
\times
\left[\frac{}{}E+\varepsilon_{k}-\Sigma_{1}(E)-\Sigma_{2}(E)\right]+
\\
\displaystyle
+\left[\frac{}{}1-\partial_{E}\Sigma_{1}(E)-\partial_{E}\Sigma_{2}(E)\right]\times
\\
\displaystyle
\times
\left[\frac{}{}E-\varepsilon_{k}-\Sigma_{1}(E)+\Sigma_{2}(E)\right]-
\\
\displaystyle
-8\mid\frac{}{}V_{k}\mid^{2}\Sigma^{\uparrow\downarrow}(E)\partial_{E}\Sigma^{\uparrow\downarrow}(E).
\end{array}
\end{equation}
In this paper, we restrict ourselves to an approximation quadratic in the fluctuating potential for 
the self-energy part. In this approximation, explicit expressions for the derivatives of its components 
with respect to energy have the form: 
\begin{equation}
\label{eq:B5}
\begin{array}{c}
\displaystyle
\partial_{E}\Sigma_{1}(E)=-\frac{V_{1}^{2}}{N}\sum_{\bf k}\frac{E^{2}+E^{2}_{k}}
{\left[E^{2}-E^{2}_{k}\right]^{2}},
\\
\displaystyle
\partial_{E}\Sigma_{2}(E)=\frac{V_{1}^{2}}{N}\sum_{\bf k}\frac{E^{2}-E^{2}_{k}+
2E\varepsilon_{k}}{\left[E^{2}-E^{2}_{k}\right]^{2}},
\\
\displaystyle
\partial_{E}\Sigma^{\uparrow\downarrow}(E)=\frac{V_{2}^{2}-V_{1}^{2}}{N}
\sum_{\bf k}\frac{4V\overline{\Delta}E\cos{k_{1}}}{\left[E^{2}-E^{2}_{k}\right]^{2}},
\end{array}
\end{equation}
where
\begin{equation}
\label{eq:B6}
\begin{array}{c}
\displaystyle
E_{k}=\sqrt{\frac{}{}\varepsilon_{k}^{2}+4V^{2}\overline{\Delta}^{2}\mid V_{k}\mid^{2}}\,, 
\end{array}
\end{equation}
Substituting expressions (\ref{eq:B1})-(\ref{eq:B5}) into (\ref{eq:matrix3}) and further into 
(\ref{eq:Sigma_H_T}) gives an explicit form of the integrand for $C_{1}$, which is not given here 
because of its cumbersomeness. Recall that the numerical values of the amplitude of the 
fluctuation-averaged superconducting order parameter $\overline{\Delta}$ in (\ref{eq:B1})-(\ref{eq:B6}) 
for different values of the Hamiltonian parameters and temperature are determined through a 
self-consistent procedure in the functional-integral method.

Note that $A^{y}$, $B^{y}$, $C^{y}$ and $D^{y}$ are proportional to $\sin{\alpha}$. Hence the topological 
index  $C_{1}$ is also proportional to $\sin{\alpha}$. This results in the fact that the superconducting 
phase with generalized $s$ symmetry ($\alpha=0$) is always topologically trivial $C_{1}=0$.


\begin{thebibliography}{00}


\bibitem{Valkov_2022} V. V. Val'kov, M. S. Shustin, S. V. Aksenov, A. O. Zlotnikov, A. D. Fedoseev, 
V. A. Mitskanm, M. Yu. Kagan, Topological superconductivity and Majorana states in low-dimensional 
systems, Phys. Usp. {\bf 65}(1), 2 (2022).

\bibitem{Nayak_2008} C. Nayak, S. H. Simon, A. Stern, M. Freedman and S. D. Sarma, Non-Abelian 
anyons and topological quantum computation, Rev. Mod. Phys. {\bf 80}(3), 1083 (2008). 

\bibitem{Zeng_2018} Bei Zeng, Xie Chen, Duan-Lu Zhou, Xiao-Gang Wen, Quantum Information Meets Quantum 
Matter From Quantum Entanglement to Topological Phase in Many-Body Systems (Springer, Berlin, 2019), 
pp. 1-364.

\bibitem{Yuto_2020} A. Yuto, G. Zongping and U. Masahito, Non-Hermitian physics, Adv. Phys. {\bf 69}(3), 
249 (2020).

\bibitem{Kawabata_2019} K. Kawabata, S. Higashikawa, Z. Gong, Y. Ashida and M. Ueda, 
Topological unification of symmetries in non-Hermitian physics, Nat. Commun. {\bf 10}(297), 1 (2019). 

\bibitem{Long_2022}  Y. Long, H. Xue, and B. Zhang, Non-Hermitian topological systems with eigenvalues 
that are always real, Phys. Rev. B {\bf 105}, L100102 (2022). 

\bibitem{Markov_2021} A. A. Markov and A. N. Rubtsov, Local marker for interacting topological insulators, 
Phys. Rev. B {\bf 104}, L081105 (2021). 

\bibitem{Groshev_2021} A. G. Groshev  and A. K. Arzhnikov, Thermal fluctuations in superconducting phases 
with chiral $d + id$ and $s$ symmetry on a triangular lattice, J. Phys.: Cond. Matt. {\bf 33}, 215604 
(2021).

\bibitem{Zhou_2008} S. Zhou and Z. Wang, Nodal $d-id$ Pairing and Topological Phases on the Triangular 
Lattice of $Na_{x}CoO_{2}yH_{2}O$ : Evidence for an Unconventional Superconducting State, 
Phys. Rev. Lett. {\bf 100}, 217002 (2008).

\bibitem{Valkov_2019} V. V. Val'kov, V. A. Mitskan, A. O. Zlotnikov, M. S. Shustin, S. V. Aksenov, 
Occurrence of Topologically Nontrivial Phases, Cascade  of Quantum Transitions, and Identification of 
Majorana Modes in Chiral Superconductors and Nanowires (Scientific Summary), J. Exp. Theor. Phys. Lett. 
{\bf 110}, 140 (2019).

\bibitem{Ishikawa_1987} K. Ishikawa and T. Matsuyama, A microscopic theory of the quantum Hall effect, 
Nuc. Phys. B {\bf 280}, 523 (1987).

\bibitem{Rachel_2010} S. Rachel and K. Le Hur, Topological insulators and Mott physics from the Hubbard 
interaction, Phys. Rev. B {\bf 82}, 075106 (2010).

\bibitem{Raghu_2008} S. Raghu, L. Qi, C. Honerkamp and S. C. Zhang, Topological Mott Insulators, 
Phys. Rev. Lett. {\bf 100}, 156401 (2008).

\bibitem{Niu_1985} Q. Niu, D. J. Thouless and Yong-Shi Wu, Quantized Hall conductance as a topological 
invariant, Phys. Rev. B {\bf 31}(6), 3372 (1985).

\bibitem{Wang_2011} L. Wang, X. Dai and  X. C. Xie, Frequency domain winding number and interaction 
effect on topological insulators, Phys. Rev. B {\bf 84}, 205116 (2011).

\bibitem{Budich_2012} B. Jan Carl, T. Ronny, L. Gang, L. Manuel and Z. Shou-Cheng, Fluctuation-induced 
topological quantum phase transitions in quantum spin-Hand anomalous-Hall insulators, Phys. Rev. B {\bf 86}, 
201407(R) (2012).

\bibitem{Zheng_2019} Z. Jun-Hui, Q. Tao and H. Walter, Interaction-enhanced integer quantum Hall effect 
in disordered systems, Phys. Rev. B {\bf 99},125138 (2019).

\bibitem{Groshev_2022} A. G. Groshev, A. K. Arzhnikov, Formation of Self-Energy Singularities by Thermal 
Fluctuations of the Superconducting Order Parameter, J. Exp. Theor. Phys. {\bf 134}(3), 305 (2022).

\bibitem{Thouless_1982} D. J. Thouless, M. Kohmoto, M. P. Nightingale and M. den Nijs, Quantized Hall 
Conductance in a Two-Dimensional Periodic Potential, Phys. Rev. Lett. {\bf 49}, 405 (1982).

\bibitem{Mahan_2000} P. G. Mahan, Many-Particle Physics (Kluwer Academic/ Plenum Publishers, New York, 2000), 
pp. 1-785.

\bibitem{Izyumov_1988} Yu. A. Izyumov and Yu. N. Skryabin, Statistical Mechanics of 
Magnetically Ordered Systems ( New York: Springer-Verlag) (1988).

\bibitem{Groshev_2020} A. G. Groshev and A. K. Arzhnikov, Self-Consistent Consideration of Fluctuations in 
Singlet Superconducting Phases with $s$ and $d$ Symmetry, J. Exp. Theor. Phys. {\bf 130}, 247 (2020).

\bibitem{Wang_2012} L. Wang, H. Jiang, Xi Dai and X. C. Xie, Phys. Rev. B {\bf 85}, 235135 (2012). 

\bibitem{Izyumov_1965} Yu. A. Izyumov, Advan es in Physics 14(56), 569 (1965).

\end{thebibliography}
\end{document}